\documentclass[utf8]{frontiersSCNS} 

\usepackage{url,hyperref,lineno,microtype,subcaption}
\usepackage[onehalfspacing]{setspace}
\def\keyFont{\fontsize{8}{11}\helveticabold }
\def\firstAuthorLast{Kiashemshaki {et~al.}} 
\def\Authors{Maryam Kiashemshaki\,$^{1}$, Zhiren Huang\,$^{1}$ and Jari Saramäki\,$^{1,2,*}$}
\def\Address{$^{1}$Department of Computer Science, Aalto University, Espoo, Finland \\
$^{2}$Helsinki Institute of Information Technology HIIT, Aalto University, Espoo, Finland}
\def\corrAuthor{Jari Saramäki}
\def\corrEmail{jari.saramaki@aalto.fi}

\begin{document}
\onecolumn
\firstpage{1}

\title[Mobility signatures]
{Mobility signatures: a tool for characterizing cities using  intercity mobility flows}

\author[\firstAuthorLast ]{\Authors} 
\address{} 
\correspondance{} 
\extraAuth{}

\maketitle
\begin{abstract}
\section{}

Understanding the patterns of human mobility between cities has various applications from transport engineering to spatial modeling of the spreading of contagious diseases. We adopt a city-centric, data-driven perspective to quantify such patterns and introduce \emph{the mobility signature} as a tool for understanding how a city (or a region) is embedded in the wider mobility network. We demonstrate the potential of the mobility signature approach through two applications that build on mobile-phone-based data from Finland. First, we use mobility signatures to show that the well-known radiation model is more accurate for mobility flows associated with larger cities, while the traditional gravity model appears a better fit for less populated areas. Second, we illustrate how the SARS-CoV-2 pandemic disrupted the mobility patterns in Finland in the spring of 2020. These two cases demonstrate the ability of the mobility signatures to quickly capture features of mobility flows that are harder to extract using more traditional methods. 
\tiny
 \keyFont{ \section{Keywords:} Travel patterns, Collective human mobility, Mobility signature, Mobile phones, OD matrix, Covid-19} 
\end{abstract}

\section{Introduction}

Collective human mobility patterns describe population movements between regions. To predict or quantify the flow volumes between regions, several classical theoretical models have been proposed by considering the impact of distances \citep{Zipf1946, Wilson1971} or intervening opportunities \citep{stouffer1940intervening, simini2012universal, yan2017universal}. From the network science perspective, the collective human mobility pattern is usually represented as a weighted mobility network \citep{barbosa2018human}. Such networks have proven to be useful for transport engineering \citep{wang2012understanding, Ren2014, guirao2018labour} and they have provided crucial information for emergency management \citep{Lu2012, huang2018mobility}. Mobility flows have been used for clustering cities \citep{ratti2010redrawing, liu2014uncovering, sen2019identifying, louail2015uncovering} and they have been shown to correlate with the socioeconomic status of cities \citep{amini2014impact, barbosa2021uncovering}. The mobility network also plays an important role in predicting the spreading of epidemics \citep{brockmann2013hidden, oliver2020mobile}, and evaluating the effects of interventions \citep{arenas2020modeling, kraemer2020effect}.

 There is, however, still a gap in quantifying the mobility characteristics of individual cities or regions. Here, a city-centric viewpoint that focuses on the mobility flows surrounding a city can be useful. As an example, mobility patterns to and from cities of different sizes and population density might respond differently to the SARS-CoV-2 pandemic.

In this paper, we introduce the city-centric mobility network similar to egocentric networks that are used in social network analysis \citep{saramaki2014persistence, heydari2018multichannel}. City-centric networks can be used to study and quantify the mobility patterns which characterize how cities are embedded in country-wide networks of mobility. In this network, a city is connected to all other cities to which there is outgoing travel, with flows to these different destinations as link weights. Normalizing the flows by the total outgoing travel flow and sorting the destinations from the most to the least visited yields a signature curve, similar to social signatures in \citep{saramaki2014persistence}. We describe this method in detail in section \ref{method}. The shape of the signature provides information about the mobility interaction patterns between a region and other regions: a smoother curve indicates that there is a more uniform flow of outgoing travelers to other areas, while a steeper curve indicates that  there are a few dominant destinations.

We apply the above methods to human mobility data for 310 municipalities in Finland using aggregated, anonymized mobile phone data from the teleoperator Telia. Then, we compare the structure of mobility signatures associated with empirical data to the mobility signatures extracted from mobility flows from two theoretical models (radiation model and gravity model). The results reveal that the mobility signatures of large municipalities are more compatible with the estimated mobility flows from the radiation model, while the signatures associated with the gravity model are a good estimation for mobility patterns for regions with medium or small populations. These results highlight a clear relation between mobility signatures and municipality size. 

Next, we focus on changes in the mobility signatures. To this end, we investigate the effects of the SARS-CoV-2 pandemic on the mobility patterns in Finland using the city-centric point of view. During the first phase of the SARS-CoV-2 outbreak, most countries across the globe established policies to control the spread of the disease by restricting human movements. Several studies \citep{Schlosser32883,10.3389/fdata.2021.718351} have investigated the structural changes in human movement from different perspectives in various regions. For example, authors in \citep{Schlosser32883} represent that long-distance travel in Germany has reduced, and consequently the mobility network is transformed into a more local and clustered one. In this paper, we monitor changes in the shapes of the mobility signatures of Finnish cities during the four months (February, March, April, and May) in 2019 and 2020 using mobile phone data. The results show that typically, there is a drop in the signature length, i.e., in the number of destinations. Most of this change is associated with reduced long-distance travel to destinations with lower mobility flows in the cities' pre-pandemic signatures.

\section{RESULTS} 

\subsection{Mobility signatures of cities}

\begin{figure}[htbp]
\begin{center}
\includegraphics[width=0.95\textwidth]{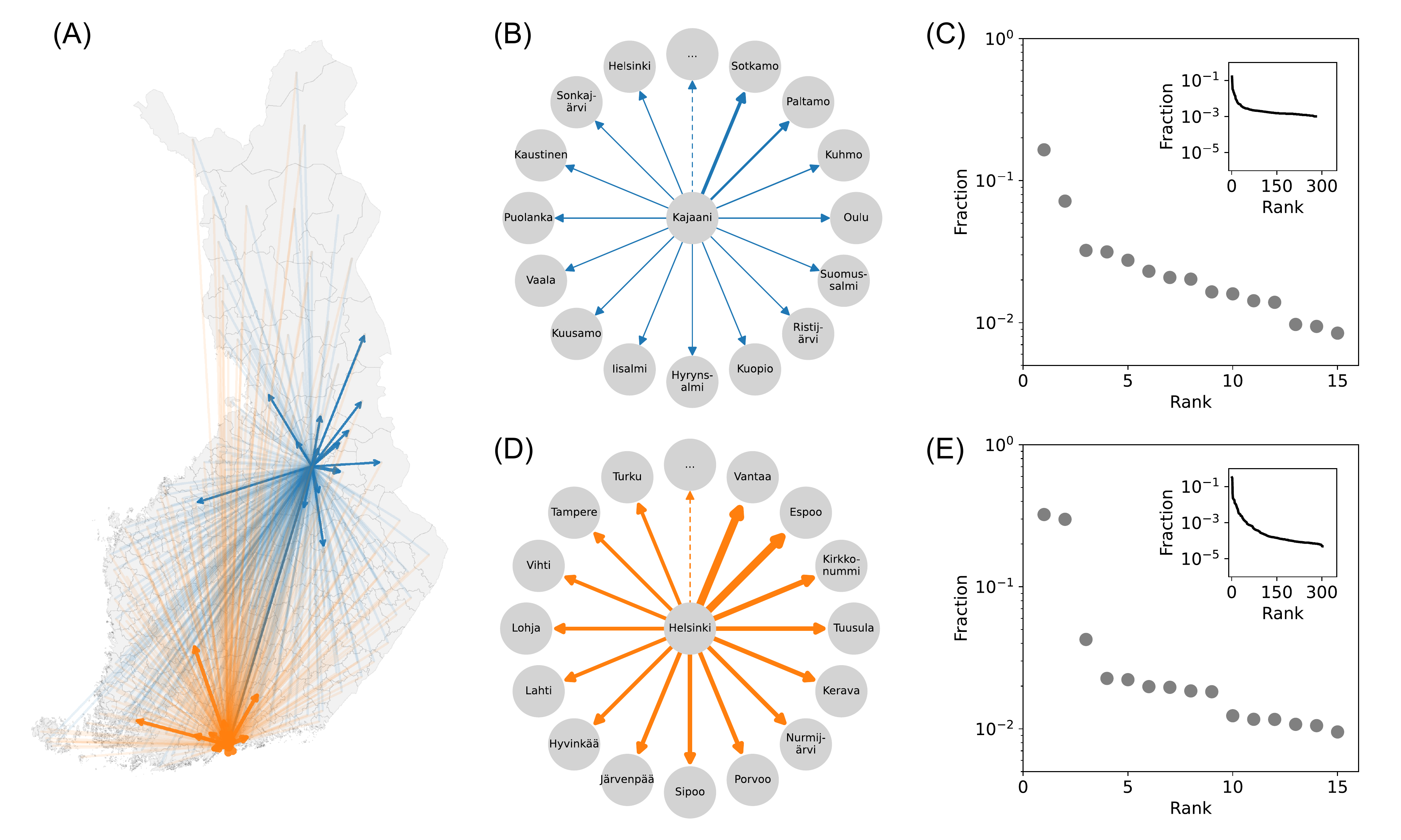}
\end{center}
\caption{City-centric networks and mobility signatures. A city-centric network studies the network surrounding one particular city/municipality. A: Outgoing mobility flows for Kajaani (blue) and Helsinki (orange). B, D: The city-centric networks for Kajaani and Helsinki, respectively. C, E: The mobility signatures for Kajaani and Helsinki; these signatures are constructed by counting the number of outgoing trips from each municipality to destinations, ranking the destinations based on this number, and then calculating the fraction of mobility flow to the total number of outgoing trips.}\label{fig:1}
\end{figure}

The mobility flow between two regions can be captured by origin-destination (OD) matrices. These matrices contain the number of trips between two geographical points or areas. There are several choices that need to be made when constructing OD matrices, including choosing spatial scales and period of observations \citep{friedrich2010generating, mungthanya2019constructing}. The spatial scale represented by the elements of the OD matrix can be, for example, a city, county, municipality, or country. While the period of observation may be limited by data availability, one can, depending on the application  study the dynamics of OD matrices by constructing daily, weekly, or monthly matrices, or alternatively aggregate over all available data.

Generally, OD  matrices can be cast as a directed weighted network, where the nodes represent chosen geographic regions, and weighted edges denote the mobility flows between them. 
In this study, we investigate the structure of country-wide mobility networks, where the nodes represent municipalities (see \ref{dataset}). Regarding the temporal dimension, in the two applications below, we study networks aggregated over an entire year as well as the dynamics of monthly networks. 

Here, we want to focus on the features of individual municipalities and cities---how are they embedded in the country-wide mobility networks? To this end, we adopt the concept of egocentric networks from social-network studies to mobility networks. An egocentric network contains information on the direct social relationships of the ego (the focal node) and her/his alters (friends, acquaintances, family members). Analogously, a city-centric network contains all mobility flows to/from a city; the country-wide mobility network can therefore be split into as many city-centric networks as there are cities. Note that in the following, we will use the terms city and city-centric network, even though the underlying data have been aggregated to the level of municipalities, as the vast majority of municipalities are technically cities.

To construct city-centric mobility networks for each municipality (Fig. \ref{fig:1} A), we use mobile-phone-based data on the number of individuals traveling from each municipality (ego) to all destinations (alters). The mobility flows are used as the link weights (Fig. \ref{fig:1} B, D). We use \emph{mobility signatures} to characterize the city-centric networks 
(Fig. \ref{fig:1} C and E). Mobility signatures are constructed as follows: we first count the number of trips to each destination (alter) and then normalize the flows by the total outgoing flow. We then sort the destinations from the most to the least visited (see Materials and Methods). In other words, we calculate the fraction of trips as a function of destination rank to generate the mobility signature of each city. 

The mobility signatures are characterized by a heavy tail as illustrated in Fig. \ref{fig:1} C and E: there is a small number of top-ranked destinations that attract a large number of trips, and numerous destinations with increasingly smaller mobility flow.

\subsection{Mobility signatures derived from theoretical models}

\begin{figure}[htbp]
\begin{center}
\includegraphics[width=0.75\textwidth]{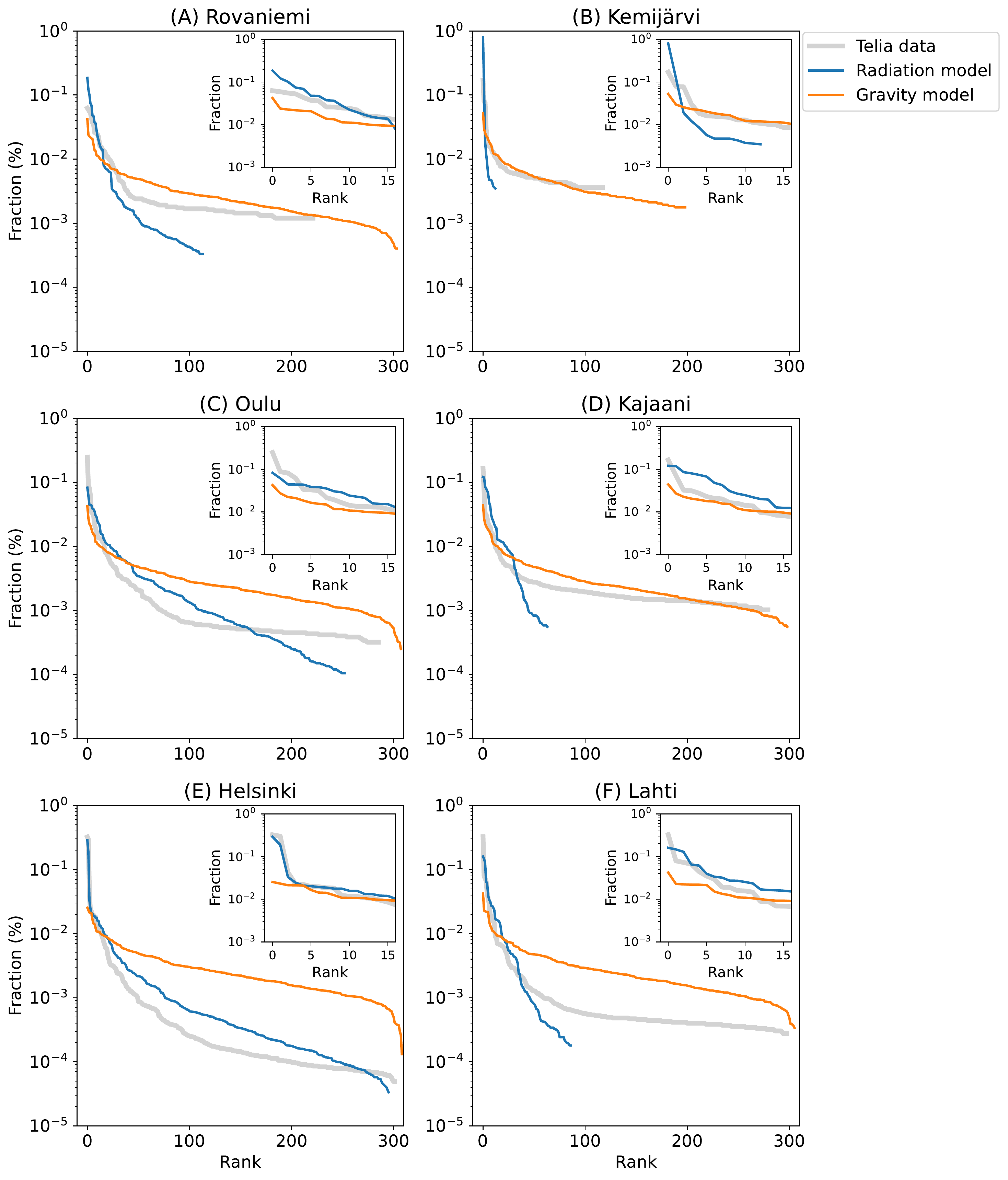}
\end{center}
\caption{Mobility signatures extracted from empirical data (mobile-phone data for 2019) and from theoretical models, for six selected cities across Finland. The two cities of the top row are in northern Finland, the cities of the middle row in central Finland, and the bottom cities in southern Finland. The first 15 top-ranked destinations are shown in the inset of each plot. Note that the cities on the left (A, C, E) are larger than the cities on the right (B, D, F).}\label{fig:2}
\end{figure}

We first explore the signatures of long-term mobility patterns and analyze the performance of two classic mobility models (i.e., radiation model and gravity model) from the perspective of signature curves, i.e., how well the theoretical models capture the mobility characteristics of individual cities. Since population data are often updated annually, these mobility flows estimated from the two models generally reflect the mobility pattern of the corresponding year. We calculate the signatures of two classic mobility models by using the population data of 2019. Note that since in Telia data, flows below 20 trips per day have been excluded due to privacy requirements, in the theoretical models, we use the same threshold and omit flows of less than 20 trips between any two cities.

Results for the mobility-derived and model-based signatures are displayed in Fig. \ref{fig:2}, for six Finnish cities. It can be seen that the radiation model performs rather well in Helsinki, especially for the top 15 destinations (Fig. 2E). For the other five cities, the signature curves from the radiation model have fewer destinations than the empirical signature curves; note that flow corresponding to less than 20 trips per day has been removed, as discussed above, which limits the signature length. The gravity model, on the other hand, generally underestimates the mobility flow for the top 15 destinations for all cities. 

\begin{figure}[htbp]
\begin{center}
\includegraphics[width=0.8\textwidth]{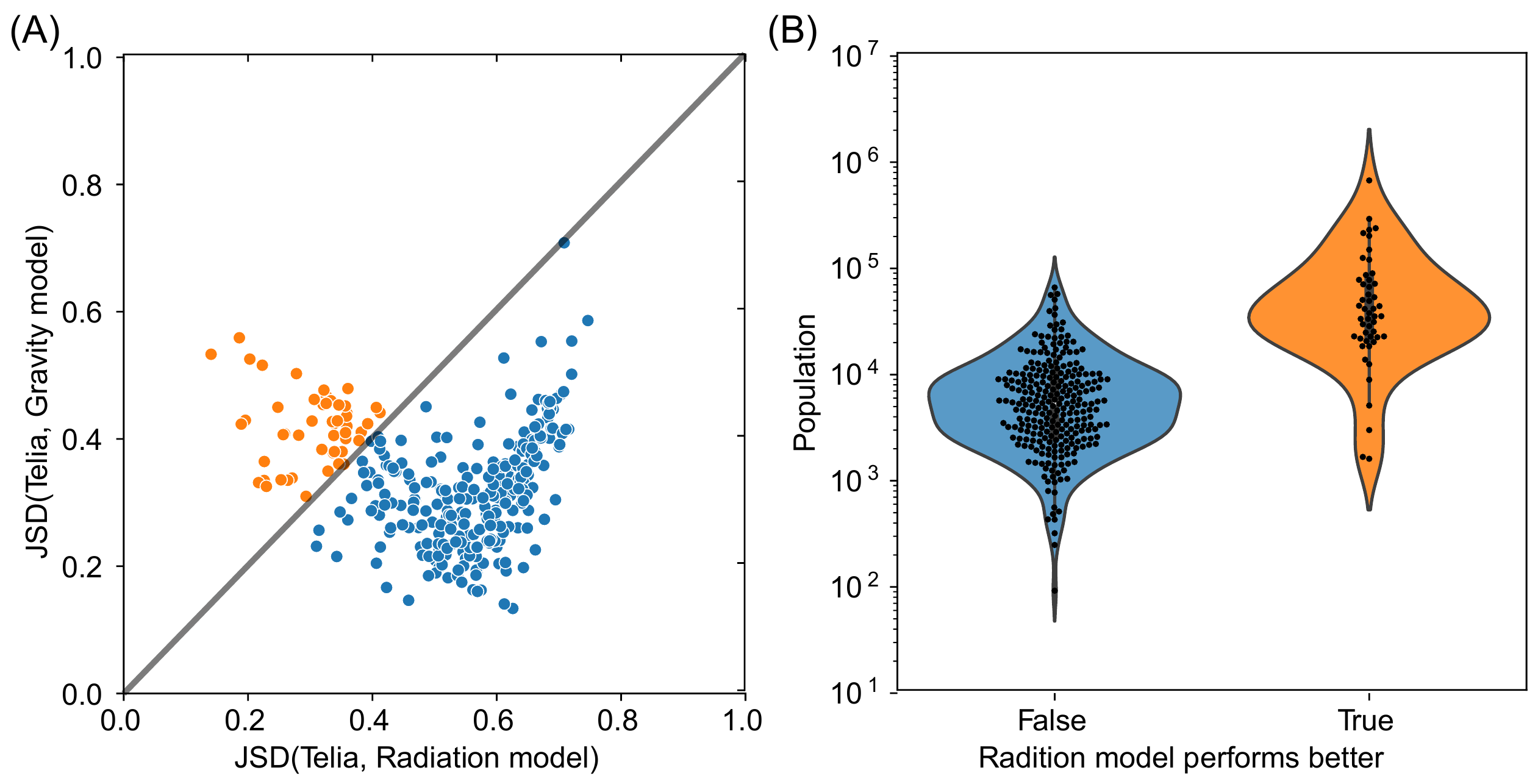}
\end{center}
\caption{A: The Jensen-Shannon Divergence (JSD) between empirical signatures and signatures from theoretical models. Orange dots represent cities whose signatures match better with the radiation model, and blue dots represent cities for which the gravity model works better. A low value of the JSD indicates a high level of similarity. B: City population distribution for the two groups. In group I (blue), the gravity model performs better than the radiation model, and the opposite is true for  group II (orange).}\label{fig:3}
\end{figure}

The above hints that the radiation model could be more accurate for larger cities (Oulu and Helsinki in Fig.\ref{fig:2}). To test this hypothesis, we need to quantify the levels of similarity between the signatures extracted from empirical observations and theoretical models. To this end, we compare the Jensen-Shannon divergence (JSD) between the mobility signatures of the empirical data and each model (see \ref{JSD}). A low value of the JSD indicates a high level of similarity. For all cities, the mean JSD of the gravity model is $0.33$, while this value for the radiation is $0.52$, indicating that the gravity model is generally more accurate. However, the city size plays a role here. As seen in Fig. \ref{fig:3}A, the gravity-model signature performs better with an average JSD of $0.31$ for almost $84\%$ of cities; these are typically small- to medium-sized cities (\ref{fig:3}B). The $16\%$ of cities whose signatures match better with the radiation model are located in the larger end of the Finnish city size distribution. 

\begin{figure}[htbp]
\begin{center}
\includegraphics[width=0.85\textwidth]{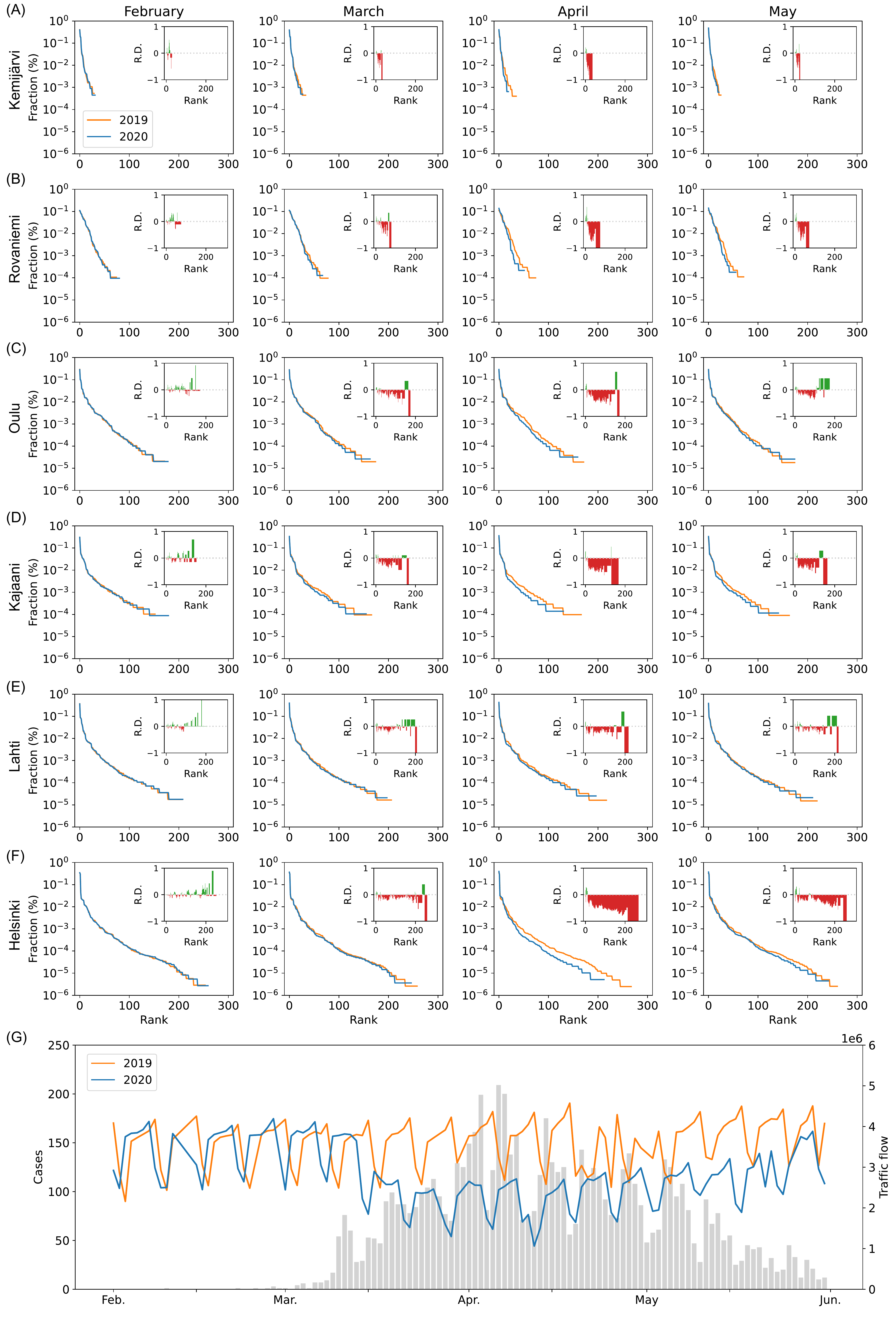}
\end{center}
\caption{A-F: Month-to-month comparison of mobility signatures during February-May in 2019 (orange curves) and 2020 (blue curves). The relative differences (i.e., R.D. in insets) are the difference between fractions of trips during 2019 and fractions of trips during 2020 divided by fractions of trips during 2019. G: The number of infected SARS-CoV-2 cases in Finland during the same time period.} \label{fig:4}
\end{figure}

\subsection{The change of mobility patterns in Finland due to SARS-CoV-2 lockdown} \label{result_3}

During the SARS-CoV-2 pandemic, countries across the globe implemented different mobility restrictions to control disease spreading. In Finland, in the spring of 2020, these policies included the closure of the nations' borders, restrictions on public activities, school closures, and travel bans in the Uusimaa region \citep{FinnishGovernmen}. There was no general lockdown, so beyond travel to and from Uusimaa, changes in mobility patterns reflect peoples' voluntary responses. To explore changes in mobility patterns during the pandemic, we study how mobility signatures have changed considering 2019 as the baseline period and compare the mobility patterns with the corresponding months in 2020. 
In addition to mobility signatures that are based on the relative number of trips, we will also look at trip distances. 

To investigate relative changes in trip numbers, we plot the mobility signatures for the same cities as above in Finland for four months in 2019 and during the first pandemic spring of 2020, as seen in Fig. \ref{fig:4} A-F. For all cities, the distance between the 2020 and 2019 signature curves is much larger for April than for the other months, which indicates that there was a disruption in mobility patterns in April 2020.
This disruption is visible as the shortening of mobility signatures, which indicates a drop in the number of traveled destinations for individual cities.

\begin{figure}[b]
\begin{center}
\includegraphics[width=\textwidth]{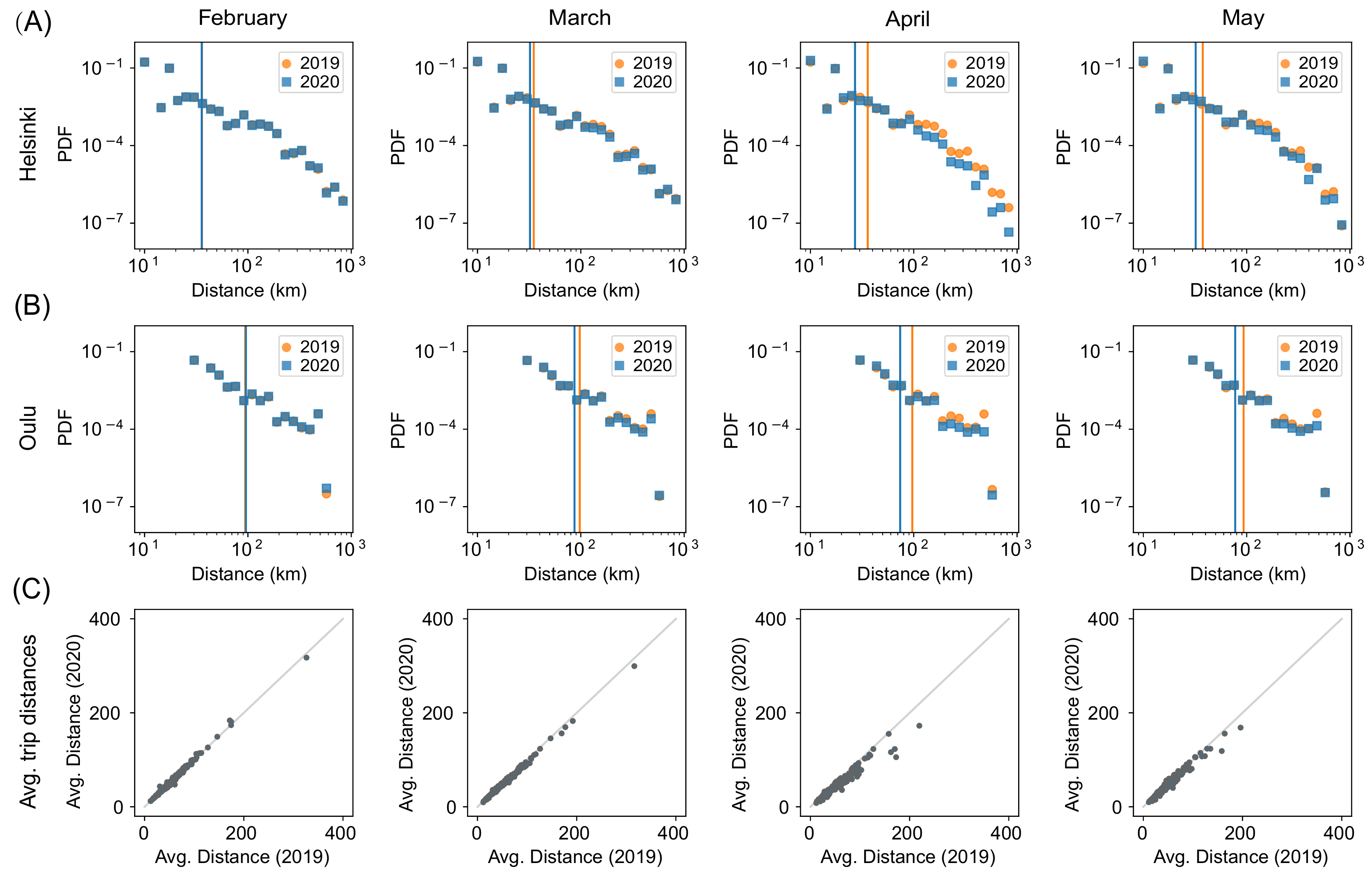}
\end{center}
\caption{Mobility changes in Finland in terms of travel distance during 2020. A: The distance distribution of outgoing trips for Helsinki; B: The same distribution for Oulu. The greatest reduction occurs for long-distance trips. The vertical lines indicate the average outgoing trip distance for each period (orange: 2019, blue: 2020). C: Scatterplot of the average outgoing trip distances for all cities for 2019 and 2020.}\label{fig:5}
\end{figure}

It also reflects the drop in the total mobility flow that coincides with the high number of infected cases in April (Fig. \ref{fig:4} G). 
Although government-imposed travel restrictions applied to the Uusimaa region only, here reflected only in the mobility signature of Helsinki (panel F), there are changes in the signatures of all six cities. 

To study how the travel distances have changed, we plot their probability density function (PDF) for the two cities of Helsinki and Oulu (Fig. \ref{fig:5} A, B). We use geographic centroids of municipalities for computing the distances.
The trip distance distributions show that for both cities, the share of long-distance trips reduced in 2020. To compare the average traveling distances for all cities during 2019 and 2020, we calculate their average outgoing trip distances. As Fig. \ref{fig:5} C illustrates, the reduction in long-distance travel is clearly larger than in shorter trips. 

Combining the above results with the results shown in Fig.\ref{fig:4}, the picture that emerges is that the drop in mobility signature length and the changes in the shapes of the signatures' tails are mainly due to reduced long-distance travel. Even though the total travel volume decreased for all trips in spring 2020, the overall pattern of short-distance travel remained mostly similar.

\section{DISCUSSION}

In this study, we introduce mobility signatures as a tool for quantifying the patterns of travel of individual cities in the overall mobility network. Similar to the signatures of egocentric networks in the social science literature, mobility signatures capture the relative importance of different destinations, and provide a way of measuring the heterogeneity of destinations.

We apply this method to mobility networks constructed from mobile phone data for 310 Finnish municipalities. To assess to what extent the mobility signatures of cities can be explained by mobility patterns generated with the two commonly-used theoretical models, we measure the similarity of the mobility signatures extracted from empirical data and models. We find that the mobility signatures generated by the radiation model match better with the empirical signatures of larger cities, whereas the gravity model is more accurate for small and mid-sized cities. 

It would be interesting to see if the above result holds for other countries as well; it is worth noting that Finland's largest cities are fairly small compared to many countries, and that in Finland the population density is generally low and the distances large (5.5 million people in a country of 338 000 km$^2$). As the radiation model is based on the concept of intervening opportunities, sparsely populated regions may be outside its scope. On the other hand, the gravity model only depends on the distance between the two cities and their populations, which may explain why it works better for smaller cities. 

Application of the mobility signature approach to the comparison of mobility patterns between 2019 and 2020 showed that the disruption caused by the SARS-CoV-2 pandemic is seen as changes in signature shapes. These changes are associated with larger reductions in long-distance travel. This observation is similar to the one made in Ref.~\cite{Schlosser32883} for Germany. For mitigating the effects of the pandemic, a reduction in long-distance travel is rather beneficial: it makes the spread of the pathogen more localized and therefore  more controllable. The difference to Germany is that the changes in mobility patterns in Finland were only partially attributable to government interventions: travel to and from the Uusimaa region in southern Finland, including the capital city of Helsinki, was restricted in April 2020, but outside this cordon sanitaire, no mobility restrictions were in place. Despite this, long-distance travel was reduced in the whole country.

An interesting future direction would be to use mobility signatures and their features as correlates in social and regional studies, e.g., by grouping together cities with similar signatures and analyzing their demographic patterns. One can also see if it is possible to perceive urban development dynamics through changes in city signatures, e.g., by combining city signatures with socioeconomic and urban development indicators.

\section{MATERIALS AND METHODS}\label{method}

\subsection{Dataset}\label{dataset}

We use aggregated and anonymized OD data provided by the mobile network operator \emph{Telia} \citep{TeliaCrowdInsights}. These OD data provide the numbers of daily trips between 310 Finnish municipalities. The number of trips has been extrapolated by the operator based on its market share to represent the whole population (Telia's market share in Finland $\sim$32\% \citep{Telia2019}). Mobility flows smaller than 20 people per day have been removed from the data for privacy. In our analysis, the intra-municipality ﬂows are discarded (i.e., we set $T_{ii}:=0$). The data are from February, March, April, and May in 2019 and 2020. The daily average number of trips between all pairs of locations is $3.4$ million (2019) and $2.7$ million (2020). For the two theoretical mobility models, we also use the population data for 2019 from \emph{Statistics Finland} \citep{paavo}.

\subsection{Computing mobility signatures}\label{analyze}

In this study, human mobility patterns are studied by focusing on cities and their connections to the rest of the country-wide mobility network. For each city, we construct a city-centric mobility network, where the city is connected to all other cities if there is outgoing travel. Link weights represent the mobility flows to different destinations. Normalizing the flows by the total outgoing flow and sorting the destinations from the most to the least visited yields a city-specific signature curve, similar to the social network studies in \citep{saramaki2014persistence}. The signature of city $i$ is defined as
\begin{equation}
\sigma_i = [\frac{T_{i1}}{T_i},\frac{T_{i2}}{T_i},...,\frac{T_{ik_i}}{T_i}],\label{eq:signature_eq}
\end{equation}

 \noindent where $T_i$ is the total number of outgoing travelers from the focal city $i$, $T_{ij}$ is the number of trips from $i$ to destination $j$, and $k_i$ is the total number of destinations with trips from $i$.
 
\subsection{Theoretical models of collective mobility}

Theoretical collective mobility models provide estimates of the number of trips between regions or cities. Theoretical models are useful in cases where knowledge of mobility flows is needed but empirical data is inaccessible, for example when modeling geographic patterns of the spread of disease without data on people's mobility. In this study, we use two well-known mobility models to investigate the mobility signatures associated with the estimated mobility flows. 

\emph{The gravity model} is inspired by Newton’s law of gravity \citep{Zipf1946}. This model determines the mobility flow between two cities from their populations and their distance. In the model, the mobility flow between two cities $i$ and $j$ with  populations of $m_i$ and $m_j$ is proportional to the product of their populations and inversely proportional to some function of their distance $f(d_{ij})$: \begin{equation}
     \label{eq:GravityDestinationDistribution}
     T_{ij}
     \propto
     \begin{cases}
     0, &\quad j=i, \\
     \frac{m_i m_j}{f(d_{ij})}, &\quad j \ne i,
     \end{cases}
    \end{equation}
    where $m_{i}$ is the number of residents who live in city $i$, $m_j$ is number of residents in city $j$ and $d_{ij}$ is the geographic distance between city $i$ and city $j$. In the original formulation of \citep{Zipf1946}, $f(d_{ij})=d_{ij}$.

    The basic version of the gravity model is evidently a gross simplification, and it has been shown that often, the model does not match with actual empirical observations \citep{masucci2013gravity, Lenormand2016}. Therefore, many researchers have investigated different distance functions \citep{flowerdew1982method} and developed the model by applying constraints to the basic gravity model \citep{Wilson1971}. In this study, we chose the model that has performed best in the literature: the doubly constrained gravity model, with a distance function that decays exponentially with distance at a rate defined by the parameter $\beta$:
    
    \begin{equation}
     \label{eq:exp}
     f(d_{ij}) =\exp{(-\beta d_{ij}).}
    \end{equation}

    The double constraint includes outgoing trips ("production") and incoming trips ("attraction") \cite{Wilson1971}. The production constraint ensures that the tot<al number of estimated trips that depart from the city $i$ equals the empirical number of total outgoing trips $n_{i+}$, and the attraction constraint is similar but regarding incoming trips:
    
    \begin{equation}
     \label{eq:GravityProductionConstrained}
     \sum_{j}T_{ij} = n_{i+},
    \end{equation}

    \begin{equation}
     \label{eq:GravityAttractionConstrained}
     \sum_{i}T_{ij} = n_{+j}.
    \end{equation}
        
The parameters of this model are fitted using the open-source library \emph{scikit-mobility} \citep{pappalardo2019scikit}.

\emph{The radiation model} \citep{simini2012universal, Ren2014} is based on the concept of spatially distributed \emph{opportunities}. Each opportunity at every location is assigned a fitness value, and the model sorts opportunities according to their distances from the origin. A traveler then chooses the closest opportunity with fitness higher than the traveler’s fitness threshold. The model predicts the commuting flows between cities $i$ and $j$ with the number of opportunities $m_i$ and $m_j$, respectively, which are a distance $d_{ij}$ apart by:
        \begin{equation}
            T_{ij} = T_i \frac{1}{1-m_i/M}\frac{m_im_j}{(m_i+s_{ij})(m_i+m_j+s_{ij})},\label{eq:radiation_eq}
        \end{equation}
        
    \noindent where $s_{ij}$ is the total number of opportunities in the circle of radius $d_{ij}$ centered at $i$, $T_i$ is the total number of travelers that start their trips from city $i$, and $M=\sum m_i$is the total number of opportunities\citep{masucci2013gravity}.  In this study, the population of each location is considered as the number of opportunities, similar to the original version of the radiation model. 

\subsection{Quantifying the similarity of pairs of city signatures}\label{JSD}

The similarity of signatures of different cities and time intervals can be investigated using the Jensen-Shannon divergence (JSD). This measure is a generalized version of the Kullback-Leibler Divergence (KLD). Unlike the KLD, the JSD is symmetric, always well defined, and bounded \citep{lin1991divergence}. The JSD of two distributions $\sigma_1$ and $\sigma_2$ is defined as
\begin{equation}
    JSD(\sigma_1,\sigma_2) = H(\frac{1}{2}\sigma_1+\frac{1}{2}\sigma_2)-\frac{1}{2} (H(\sigma_1)+H(\sigma_2)),\label{eq:jsd_eq}
\end{equation}

\noindent  where $H(\sigma_i)$ is the Shannon entropy of $\sigma_i$:

\begin{equation}
    H(\sigma_i) = - \sum_{j =1}^k T_{ij}\log{T_{ij}},\label{eq:entropy_eq}
\end{equation}

\noindent where $k$ is the total number of destinations and $T_{ij}$ represents the number of outgoing trips from city $i$ to city $j$. The JSD values are between 0 and 1, so that the closer the value is to zero, the more similar the pair of signatures is. 

\section*{Conflict of Interest Statement}

The authors declare that they have no competing interests.

\section*{Data Availability Statement}

The mobile phone data used in this article will not be shared because of the confidentiality agreement with the data provider.

\section*{Author Contributions}

J.S., M.K., and Z.H. designed the research; M.K. and Z.H. processed the data and performed the analysis; M.K., Z.H., and J.S. analyzed the results and wrote the manuscript; All authors read, commented, and approved the final version of the manuscript.

\section*{Funding}

The study is part of the NetResilience consortium funded by the Strategic Research Council at the Academy of Finland (grant numbers 345188 and 345183).

\section*{Acknowledgments}

We thank Sara Heydari for the discussions and comments on the manuscript. The calculations presented above were performed using computer resources within the Aalto University School of Science "Science-IT" project.
 
\bibliographystyle{frontiersinSCNS_ENG_HUMS}
\bibliography{ref}

\begin{thebibliography}{36}
\providecommand{\natexlab}[1]{#1}
\expandafter\ifx\csname urlstyle\endcsname\relax
  \providecommand{\doi}[1]{doi:\discretionary{}{}{}#1}\else
  \providecommand{\doi}{doi:\discretionary{}{}{}\begingroup
  \urlstyle{rm}\Url}\fi
\providecommand{\selectlanguage}[1]{\relax}
\providecommand{\bibAnnoteFile}[1]{%
  \IfFileExists{#1}{\begin{quotation}\noindent\textsc{Key:} #1\\
  \textsc{Annotation:}\ \input{#1}\end{quotation}}{}}
\providecommand{\bibAnnote}[2]{%
  \begin{quotation}\noindent\textsc{Key:} #1\\
  \textsc{Annotation:}\ #2\end{quotation}}

\bibitem[{Amini et~al.(2014)Amini, Kung, Kang, Sobolevsky, and
  Ratti}]{amini2014impact}
Amini, A., Kung, K., Kang, C., Sobolevsky, S., and Ratti, C. (2014).
\newblock The impact of social segregation on human mobility in developing and
  industrialized regions.
\newblock \emph{EPJ Data Science} 3, 1--20
\bibAnnoteFile{amini2014impact}

\bibitem[{Arenas et~al.(2020)Arenas, Cota, G{\'o}mez-Garde{\~n}es, G{\'o}mez,
  Granell, Matamalas et~al.}]{arenas2020modeling}
Arenas, A., Cota, W., G{\'o}mez-Garde{\~n}es, J., G{\'o}mez, S., Granell, C.,
  Matamalas, J.~T., et~al. (2020).
\newblock Modeling the spatiotemporal epidemic spreading of covid-19 and the
  impact of mobility and social distancing interventions.
\newblock \emph{Physical Review X} 10, 041055
\bibAnnoteFile{arenas2020modeling}

\bibitem[{Barbosa et~al.(2018)Barbosa, Barthelemy, Ghoshal, James, Lenormand,
  Louail et~al.}]{barbosa2018human}
Barbosa, H., Barthelemy, M., Ghoshal, G., James, C.~R., Lenormand, M., Louail,
  T., et~al. (2018).
\newblock Human mobility: Models and applications.
\newblock \emph{Physics Reports} 734, 1--74
\bibAnnoteFile{barbosa2018human}

\bibitem[{Barbosa et~al.(2021)Barbosa, Hazarie, Dickinson, Bassolas, Frank,
  Kautz et~al.}]{barbosa2021uncovering}
Barbosa, H., Hazarie, S., Dickinson, B., Bassolas, A., Frank, A., Kautz, H.,
  et~al. (2021).
\newblock Uncovering the socioeconomic facets of human mobility.
\newblock \emph{Scientific reports} 11, 1--13
\bibAnnoteFile{barbosa2021uncovering}

\bibitem[{Brockmann and Helbing(2013)}]{brockmann2013hidden}
Brockmann, D. and Helbing, D. (2013).
\newblock The hidden geometry of complex, network-driven contagion phenomena.
\newblock \emph{Science} 342, 1337--1342
\bibAnnoteFile{brockmann2013hidden}

\bibitem[{{Finnish Government}(2021)}]{FinnishGovernmen}
[Dataset] {Finnish Government} (2021).
\newblock Movement restrictions - region of uusimaa.
\newblock [Online; accessed 12-November-2021]
\bibAnnoteFile{FinnishGovernmen}

\bibitem[{Flowerdew and Aitkin(1982)}]{flowerdew1982method}
Flowerdew, R. and Aitkin, M. (1982).
\newblock A method of fitting the gravity model based on the poisson
  distribution.
\newblock \emph{Journal of regional science} 22, 191--202
\bibAnnoteFile{flowerdew1982method}

\bibitem[{Friedrich et~al.(2010)Friedrich, Immisch, Jehlicka, Otterst{\"a}tter,
  and Schlaich}]{friedrich2010generating}
Friedrich, M., Immisch, K., Jehlicka, P., Otterst{\"a}tter, T., and Schlaich,
  J. (2010).
\newblock Generating origin--destination matrices from mobile phone
  trajectories.
\newblock \emph{Transportation research record} 2196, 93--101
\bibAnnoteFile{friedrich2010generating}

\bibitem[{Guirao et~al.(2018)Guirao, Campa, and Casado-Sanz}]{guirao2018labour}
Guirao, B., Campa, J.~L., and Casado-Sanz, N. (2018).
\newblock Labour mobility between cities and metropolitan integration: The role
  of high speed rail commuting in spain.
\newblock \emph{Cities} 78, 140--154
\bibAnnoteFile{guirao2018labour}

\bibitem[{Heydari et~al.(2018)Heydari, Roberts, Dunbar, and
  Saram{\"a}ki}]{heydari2018multichannel}
Heydari, S., Roberts, S.~G., Dunbar, R.~I., and Saram{\"a}ki, J. (2018).
\newblock Multichannel social signatures and persistent features of ego
  networks.
\newblock \emph{Applied network science} 3, 1--13
\bibAnnoteFile{heydari2018multichannel}

\bibitem[{Huang et~al.(2018)Huang, Wang, Zhang, Gao, and
  Schich}]{huang2018mobility}
Huang, Z., Wang, P., Zhang, F., Gao, J., and Schich, M. (2018).
\newblock A mobility network approach to identify and anticipate large crowd
  gatherings.
\newblock \emph{Transportation research part B: methodological} 114, 147--170
\bibAnnoteFile{huang2018mobility}

\bibitem[{Kraemer et~al.(2020)Kraemer, Yang, Gutierrez, Wu, Klein, Pigott
  et~al.}]{kraemer2020effect}
Kraemer, M.~U., Yang, C.-H., Gutierrez, B., Wu, C.-H., Klein, B., Pigott,
  D.~M., et~al. (2020).
\newblock The effect of human mobility and control measures on the covid-19
  epidemic in china.
\newblock \emph{Science} 368, 493--497
\bibAnnoteFile{kraemer2020effect}

\bibitem[{Lenormand et~al.(2016)Lenormand, Bassolas, and
  Ramasco}]{Lenormand2016}
Lenormand, M., Bassolas, A., and Ramasco, J.~J. (2016).
\newblock Systematic comparison of trip distribution laws and models.
\newblock \emph{Journal of Transport Geography} 51, 158--169
\bibAnnoteFile{Lenormand2016}

\bibitem[{Lin(1991)}]{lin1991divergence}
Lin, J. (1991).
\newblock Divergence measures based on the shannon entropy.
\newblock \emph{IEEE Transactions on Information theory} 37, 145--151
\bibAnnoteFile{lin1991divergence}

\bibitem[{Liu et~al.(2014)Liu, Sui, Kang, and Gao}]{liu2014uncovering}
Liu, Y., Sui, Z., Kang, C., and Gao, Y. (2014).
\newblock Uncovering patterns of inter-urban trip and spatial interaction from
  social media check-in data.
\newblock \emph{PloS one} 9, e86026
\bibAnnoteFile{liu2014uncovering}

\bibitem[{Louail et~al.(2015)Louail, Lenormand, Picornell, Cant{\'u}, Herranz,
  Frias-Martinez et~al.}]{louail2015uncovering}
Louail, T., Lenormand, M., Picornell, M., Cant{\'u}, O.~G., Herranz, R.,
  Frias-Martinez, E., et~al. (2015).
\newblock Uncovering the spatial structure of mobility networks.
\newblock \emph{Nature communications} 6, 1--8
\bibAnnoteFile{louail2015uncovering}

\bibitem[{Lu et~al.(2012)Lu, Bengtsson, and Holme}]{Lu2012}
Lu, X., Bengtsson, L., and Holme, P. (2012).
\newblock Predictability of population displacement after the 2010 haiti
  earthquake.
\newblock \emph{Proc. Natl. Acad. Sci. (USA)} 109, 11576--11581
\bibAnnoteFile{Lu2012}

\bibitem[{Masucci et~al.(2013)Masucci, Serras, Johansson, and
  Batty}]{masucci2013gravity}
Masucci, A.~P., Serras, J., Johansson, A., and Batty, M. (2013).
\newblock Gravity versus radiation models: On the importance of scale and
  heterogeneity in commuting flows.
\newblock \emph{Physical Review E} 88, 022812
\bibAnnoteFile{masucci2013gravity}

\bibitem[{Mungthanya et~al.(2019)Mungthanya, Phithakkitnukoon, Demissie,
  Kattan, Veloso, Bento et~al.}]{mungthanya2019constructing}
Mungthanya, W., Phithakkitnukoon, S., Demissie, M.~G., Kattan, L., Veloso, M.,
  Bento, C., et~al. (2019).
\newblock Constructing time-dependent origin-destination matrices with adaptive
  zoning scheme and measuring their similarities with taxi trajectory data.
\newblock \emph{IEEE Access} 7, 77723--77737
\bibAnnoteFile{mungthanya2019constructing}

\bibitem[{Oliver et~al.(2020)Oliver, Lepri, Sterly, Lambiotte, Deletaille,
  Nadai et~al.}]{oliver2020mobile}
Oliver, N., Lepri, B., Sterly, H., Lambiotte, R., Deletaille, S., Nadai, M.~D.,
  et~al. (2020).
\newblock Mobile phone data for informing public health actions across the
  covid-19 pandemic life cycle.
\newblock \emph{Science Advances} 6, eabc0764
\bibAnnoteFile{oliver2020mobile}

\bibitem[{Pappalardo et~al.(2019)Pappalardo, Simini, Barlacchi, and
  Pellungrini}]{pappalardo2019scikit}
Pappalardo, L., Simini, F., Barlacchi, G., and Pellungrini, R. (2019).
\newblock scikit-mobility: A python library for the analysis, generation and
  risk assessment of mobility data.
\newblock \emph{arXiv preprint arXiv:1907.07062}
\bibAnnoteFile{pappalardo2019scikit}

\bibitem[{Potgieter et~al.(2021)Potgieter, Fabris-Rotelli, Kimmie,
  Dudeni-Tlhone, Holloway, Janse~van Rensburg
  et~al.}]{10.3389/fdata.2021.718351}
Potgieter, A., Fabris-Rotelli, I.~N., Kimmie, Z., Dudeni-Tlhone, N., Holloway,
  J.~P., Janse~van Rensburg, C., et~al. (2021).
\newblock Modelling representative population mobility for covid-19 spatial
  transmission in south africa.
\newblock \emph{Frontiers in Big Data} 4, 82.
\newblock \doi{10.3389/fdata.2021.718351}
\bibAnnoteFile{10.3389/fdata.2021.718351}

\bibitem[{Ratti et~al.(2010)Ratti, Sobolevsky, Calabrese, Andris, Reades,
  Martino et~al.}]{ratti2010redrawing}
Ratti, C., Sobolevsky, S., Calabrese, F., Andris, C., Reades, J., Martino, M.,
  et~al. (2010).
\newblock Redrawing the map of great britain from a network of human
  interactions.
\newblock \emph{PloS one} 5, e14248
\bibAnnoteFile{ratti2010redrawing}

\bibitem[{Ren et~al.(2014)Ren, Ercsey-Ravasz, Wang, Gonz{\'{a}}lez, and
  Toroczkai}]{Ren2014}
Ren, Y., Ercsey-Ravasz, M., Wang, P., Gonz{\'{a}}lez, M.~C., and Toroczkai, Z.
  (2014).
\newblock {Predicting commuter flows in spatial networks using a radiation
  model based on temporal ranges}.
\newblock \emph{Nature Communications} 5.
\newblock \doi{10.1038/ncomms6347}
\bibAnnoteFile{Ren2014}

\bibitem[{Saram{\"a}ki et~al.(2014)Saram{\"a}ki, Leicht, L{\'o}pez, Roberts,
  Reed-Tsochas, and Dunbar}]{saramaki2014persistence}
Saram{\"a}ki, J., Leicht, E.~A., L{\'o}pez, E., Roberts, S.~G., Reed-Tsochas,
  F., and Dunbar, R.~I. (2014).
\newblock Persistence of social signatures in human communication.
\newblock \emph{Proceedings of the National Academy of Sciences} 111, 942--947
\bibAnnoteFile{saramaki2014persistence}

\bibitem[{Schlosser et~al.(2020)Schlosser, Maier, Jack, Hinrichs, Zachariae,
  and Brockmann}]{Schlosser32883}
Schlosser, F., Maier, B.~F., Jack, O., Hinrichs, D., Zachariae, A., and
  Brockmann, D. (2020).
\newblock Covid-19 lockdown induces disease-mitigating structural changes in
  mobility networks.
\newblock \emph{Proc. Natl. Acad. Sci. (USA)} 117, 32883--32890
\bibAnnoteFile{Schlosser32883}

\bibitem[{Sen and Dietz(2019)}]{sen2019identifying}
Sen, A. and Dietz, L.~W. (2019).
\newblock Identifying travel regions using location-based social network
  check-in data.
\newblock \emph{Frontiers in big Data} 2, 12
\bibAnnoteFile{sen2019identifying}

\bibitem[{Simini et~al.(2012)Simini, Gonz{\'a}lez, Maritan, and
  Barab{\'a}si}]{simini2012universal}
Simini, F., Gonz{\'a}lez, M.~C., Maritan, A., and Barab{\'a}si, A.-L. (2012).
\newblock A universal model for mobility and migration patterns.
\newblock \emph{Nature} 484, 96--100
\bibAnnoteFile{simini2012universal}

\bibitem[{StatFinland(2020)}]{paavo}
[Dataset] StatFinland (2020).
\newblock {Paavo postal code area statistics}.
\newblock \url{https://www.stat.fi/tup/paavo/index_en.html}.
\newblock [Online; accessed 10-Sep.-2021]
\bibAnnoteFile{paavo}

\bibitem[{Stouffer(1940)}]{stouffer1940intervening}
Stouffer, S.~A. (1940).
\newblock Intervening opportunities: a theory relating mobility and distance.
\newblock \emph{American sociological review} 5, 845--867
\bibAnnoteFile{stouffer1940intervening}

\bibitem[{Telia(2019)}]{Telia2019}
[Dataset] Telia (2019).
\newblock {Telia} markets and brands.
\newblock [Online; accessed 10-Sep.-2021]
\bibAnnoteFile{Telia2019}

\bibitem[{Telia(2020)}]{TeliaCrowdInsights}
[Dataset] Telia (2020).
\newblock Crowd insights service.
\bibAnnoteFile{TeliaCrowdInsights}

\bibitem[{Wang et~al.(2012)Wang, Hunter, Bayen, Schechtner, and
  Gonz{\'a}lez}]{wang2012understanding}
Wang, P., Hunter, T., Bayen, A.~M., Schechtner, K., and Gonz{\'a}lez, M.~C.
  (2012).
\newblock Understanding road usage patterns in urban areas.
\newblock \emph{Scientific reports} 2, 1--6
\bibAnnoteFile{wang2012understanding}

\bibitem[{Wilson(1971)}]{Wilson1971}
Wilson, A.~G. (1971).
\newblock A family of spatial interaction models, and associated developments.
\newblock \emph{Environment and Planning A} 3, 1--32
\bibAnnoteFile{Wilson1971}

\bibitem[{Yan et~al.(2017)Yan, Wang, Gao, and Lai}]{yan2017universal}
Yan, X.-Y., Wang, W.-X., Gao, Z.-Y., and Lai, Y.-C. (2017).
\newblock Universal model of individual and population mobility on diverse
  spatial scales.
\newblock \emph{Nature communications} 8, 1--9
\bibAnnoteFile{yan2017universal}

\bibitem[{Zipf(1946)}]{Zipf1946}
Zipf, G.~K. (1946).
\newblock {The P1 P2/D Hypothesis: On the Intercity Movement of Persons}.
\newblock \emph{American Sociological Review} 11, 677
\bibAnnoteFile{Zipf1946}

\end{thebibliography}
\end{document}